\documentclass{jpp}
\usepackage{graphicx}
\usepackage{epstopdf, epsfig}
\usepackage{hyperref}

\usepackage{amsfonts}
\usepackage{times}
\usepackage{graphicx}
\usepackage{natbib}

\renewcommand{\vec}[1]{\mbox{\boldmath $#1$}}

\def\dd{{\textrm d}}
\def \Om  {{\it \Omega}}
\def \rin {r_{\rm in}}
\def \mur {r_{\rm in}}
\def\Rin{R_{\rm in}}
\def\Rout{R_{\rm out}}
\def \Rey {\ensuremath{\rm{Re}}}
\def \Remax {\ensuremath{\rm{Re_{max}}}}
\def \Ha {\ensuremath{\rm{Ha}}}
\def \Hamin {\ensuremath{\rm{Ha_{min}}}}
\def \Pm {\ensuremath{\rm{Pm}}}

\def\nuT{\nu_{\rm T}}
\def\C{Chandrasekhar}
\def \d   {\rm d}

\def \Mm {\ensuremath{\rm{Mm}}}

\def\beg{\begin{equation}}
\def\ende{\end{equation}}
\newcommand{\gsim}{\lower.7ex\hbox{$\;\stackrel{\textstyle>}{\sim}\;$}}
\newcommand{\lsim}{\lower.7ex\hbox{$\;\stackrel{\textstyle<}{\sim}\;$}}
\renewcommand{\vec}[1]{\mbox{\boldmath $#1$}}
 
\def\Om{{\it \Omega}}

\def\A{{Alfv\'en}}

\title{Non-diffusive angular momentum transport  in   rotating $\bf z$-pinches \footnote{\today} }

\author{G. R\"udiger
  \corresp{\email{gruediger@aip.de}},
  M. Schultz}

\affiliation{Leibniz-Institut f\"ur Astrophysik Potsdam, An der Sternwarte 16, D-14482 Potsdam, Germany
}

\begin{document}

\maketitle

\begin{abstract}
The stability of conducting  Taylor-Couette flows    under the presence of 
toroidal  magnetic background fields is considered. For strong enough magnetic amplitudes  such magnetohydrodynamic  flows are unstable against nonaxisymmetric perturbations which may also  transport angular momentum. In accordance with  the often used  diffusion approximation one expects  the  angular momentum transport vanishing for rigid rotation. In    the sense of a nondiffusive $\Lambda$ effect, however,  even for {\em rigidly} rotating $z$-pinches 
an axisymmetric  angular momentum flux appears which is directed outward (inward) for large (small) magnetic Mach numbers. 
The internal rotation  in a magnetized rotating tank can thus never be uniform.
Those particular  rotation laws are used to estimate the value of the instability-induced eddy viscosity for which  the  nondiffusive $\Lambda$ effect and the diffusive shear-induced transport compensate each other. The results provide   the well-known  \cite{SS73} viscosity ansatz  leading to 
 numerical   values  linearly growing  with  the Reynolds number of rotation.
\end{abstract}
\section{Introduction}
A hydrodynamic Taylor-Couette flow with rotation profiles beyond the Rayleigh limit is stable against axi- and nonaxisymmetric perturbations.  It is unstable, however,  against axisymmetric perturbations under the presence of axial magnetic background fields \citep{V59,RZ01,JG01})  and it is unstable against nonaxisymmetric perturbations under the presence of azimuthal fields \citep{T57,OP96}.
For axial fields  and given magnetic Prandtl number 
\beg
\Pm = \frac{\nu}{\eta}
\label{Pm}
\ende
(with  molecular viscosity $\nu$ and  magnetic resistivity $\eta$) there exists always  a critical magnetic field amplitude   with a minimal   Reynolds number. These  numbers  are running with
$1/\Pm$ so that for the small magnetic Prandtl numbers of liquid metals  the critical
Reynolds numbers  basically exceed  values of $10^6$  \citep{RS01}. This very high Reynolds number    is the main reason that to date the standard magnetorotational instability has not yet been realized in  laboratory experiments. 

For much lower critical Reynolds numbers the Couette flow becomes unstable if the magnetic background field is toroidal or has a toroidal component. 
 \cite{HS06}  demonstrated  that all flows with negative shear and  current-free background fields are stable against axisymmetric perturbations.  There and also here   the flows are assumed as unbounded in the axial direction.    The existing instabilities must  thus  always be nonaxisymmetric. For the absolutely lowest possible magnetic field amplitude the critical
 Reynolds number for the onset of this so-called Azimuthal MagnetoRotational Instability (AMRI)
for liquid metals such as  gallium or sodium is only  O($10^3$), hence there is  a very strong reduction compared with the Reynolds numbers needed for the  magnetorotational instability with axial fields. Not surprisingly, the AMRI at and slightly beyond  the Rayleigh line has already been realized in the laboratory  \citep{SS14}. 

The nonaxisymmetric AMRI also  exists for flows  rotating with a positive shear $\dd\Om/\dd R$ (``super-rotation''). This is insofar  of   relevance as super-rotating Taylor-Couette flows are prototypes of very stable hydrodynamic flows (but  see \cite{D17} for a nonaxisymmetric instability at high Reynolds numbers). For conducting fluids with magnetic Prandtl numbers less or larger than  unity such flows can easily be destabilized with supercritical toroidal magnetic fields.  The needed  Reynolds numbers (of the outer cylinder) for flows with stationary inner cylinder are only O(100), see    \cite{SK15}; \cite{RG18}.

Another instability exists for conducting fluids wherein  axial electric currents produce azimuthal magnetic fields of radial profiles  less steep than the vacuum profile  $1/R$ where $R$ is the radius  in cylindric coordinates. 
The relation 
\beg
\frac{{\rm{d}}}{{\rm{d}}R}( R B_\phi^2) \leq 0
\label{tay}
\ende
is a sufficient and necessary condition for stability of a stationary ideal fluid against nonaxisymmetric perturbations \citep{T73}. One finds instability in particular for the azimuthal field with the radial profile $B_\phi\propto R$ produced by   a uniform electric current. The existence of a nonaxisymmetric instability for such  a (nonrotating)  $z$-pinch has been shown by \cite{SS12} using the liquid GaInSn alloy as the conducting fluid penetrated by an axial electric current of $\simeq 3$ kAmp. 

The  combination of a  current-free magnetic field $B_\phi\propto 1/R$  and the  rotation profile $\Om\propto1/R^2$ of the potential flow   belongs  to a particular class of MHD flows defined by the condition that  the magnetic Mach number $\Mm$  in the relation
\beg
\vec{U}= {\Mm}\ \vec{U}_{\rm A},
 \label{Mm}
 \ende
with $\vec U$  the flow velocity and   $\vec{U}_{\rm A}=\vec{B}/ \sqrt{\mu_0\rho}$ its \A\ speed, is a constant value \citep{C56}. Applied to Taylor-Couette flows the radial profiles of the flow velocity ${U_\phi}$ and $B_\phi$ are required as   identical.  All such flows  are stable in the absence of diffusive effects. On the other hand, it is known  that the potential flow of real fluids with $\Om\propto 1/R^2$ can easily  be destabilized by the current-free  toroidal magnetic field with $B_\phi\propto 1/R$ \citep{RHS07}. 
All these MHD flows possess marginal instabilities for Reynolds numbers as a function of Hartmann numbers where these values  do not depend on $\Pm$ for $\Pm\to 0$. Even in the inductionless approximation $\Pm=0$ these eigenvalues remain finite.  A prominent example  is also  the rigidly rotating $z$-pinch where the flow $ U_\phi$  and the field $B_\phi $are both  proportionate to  the cylinder radius $R$.

This  sort of    Taylor-Couette flows  will be considered  in the present paper to probe its qualification  to transport angular momentum. In the cylinder coordinates used for Taylor-Couette flows (unbounded in $z$) only the radial component $T_R$ of the total  stress tensor 
must be considered which is formed by the ($R\phi$) components of the difference of Reynolds stress and Maxwell stress.
In  the so-called  diffusion  approximation  the $T_R$ component has been written as  
\beg
T_R=-\nu_{\rm T}  R\frac{\dd\Om}{\dd R}, 
\label{Bou}
\ende
with positive  eddy viscosity $\nuT$. Such a relation   has originally  been formulated for  hydrodynamical turbulence  \citep{B97} based on the observation that in  a rigidly rotating fluid no angular momentum is transported.
We shall see, however,  that for the pinch-type    instability the diffusion approximation (\ref{Bou}) does not  hold  for uniform rotation. This gives a new possibility to estimate the magnetic-instability-induced turbulent viscosity for differentially rotating magnetized containers. The resulting expression can also be considered as a confirmation of the viscosity approximation  introduced by  \cite{SS73} and \cite{P81} to the accretion disk theory as an  explanation of the angular momentum transport in thin  disks. 

Furthermore, some   important applications  of stellar physics are based on the Eq. (\ref{Bou}).
We know from helioseismology that  the solar radiative core  rotates  rigidly.  One needs effective 
viscosities of $10^4$ times the molecular  value  to explain the decay of an initial rotation law  within the lifetime of the Sun. 
Also the  Maxwell stress theory of this decay   needs an increase of the microscopic viscosity by a few orders of magnitude  for the explanation of the rigid rotation \citep{CM92,RK96}. It remains   to test whether the angular momentum transport 
by magnetic instabilities of fossil internal  toroidal fields  is strong enough to produce the quasi-uniform inner rotation of the Sun.

The lithium at the surface of cool main-sequence stars decays with a timescale of 1 Gyr. It is burned at temperatures in excess of $2.6 \times 10^6$ K, 
which exists about 40.000 km below the base of the solar convection zone. There must be a diffusion process  down to this layer with 
the burning temperature. A slow transport process is needed which is only one or two orders of magnitude faster than the 
molecular diffusion. The molecular diffusion beneath the solar convection zone must be increased but only to about $10^3$ cm$^2$/s. 

In this paper we present the linear theory of rigidly rotating $z$-pinches (with homogeneous axial electric current)   for  two sorts of boundary conditions  in order to calculate the instability-induced normalized radial transport of angular momentum.
The basic  equations of magnetohydrodynamics (MHD) are presented   in Sect. \ref{Equations}. The eigenfunctions  of the unstable solutions  and the instability-originated angular momentum transport (its ``$\Lambda$ effect'') for given magnetic Prandtl number   are presented in Sect.~\ref{Rotating}. The transition to differential rotation also in terms of an eddy viscosity is discussed in  Sect.  \ref{Nonuniform} while Sect. \ref{Conclusions}  contains a short discussion of the results. 
\section{The Equations}\label{Equations}
The equations of the magnetic-instability theory are the well-known MHD equations
\begin{eqnarray}
 \frac{\partial \vec{U}}{\partial t} + (\vec{U}\cdot \nabla)\vec{U}& =& -\frac{1}{\rho} \nabla P + \nu \Delta \vec{U} 
   + \frac{1}{\mu_0\rho}{\textrm{curl}}\vec{B} \times \vec{B},\nonumber\\
 \frac{\partial \vec{B}}{\partial t}&=& {\textrm{curl}} (\vec{U} \times \vec{B}) + \eta \Delta\vec{B}  
   \label{mhd2}
\end{eqnarray}
with $  {\textrm{div}}\ \vec{U} = {\textrm{div}}\ \vec{B} = 0$ for an incompressible fluid.
$\vec{U}$ is the velocity vector, $\vec{B}$ the magnetic field vector and $P$ the pressure. The  basic state in the cylindric system with the
coordinates ($R,\phi,z$) is \mbox{$ U_R=U_z=B_R=B_z=0$} for the poloidal components and 
\beg
\Om  = a  + \frac{b}{R^2}
\label{Om}
\ende
 for the rotation law with the constants 
$ a=\Om_{\rm in}(\mu-\mur^2)/(1-\mur^2)$ and 
$ b= \Om_{\rm in} R_{\rm in}^2(1-\mu)/(1-\mur^2)$.
 Here $\rin={R_{\rm in}}/{R_{\rm out}}$ is the ratio of the inner cylinder radius $R_{\rm in}$ and the outer cylinder 
radius $R_{\rm out}$.  $\Om_{\rm in}$ and $\Om_{\rm out}$ are the angular velocities of the inner and outer cylinders, 
respectively. With the definition 
\beg
\mu=\frac{ \Om_{\rm out}}{\Om_{\rm in}}
\label{mu}
\ende
$\mu=1$ describes solid-body  rotation with uniform $\Om$ while $\mu<1$ belongs to rotation laws with negative radial shear (``sub-rotation''). Super-rotation (positive shear, ${\rm d} \Om/{\rm d} R>0$) leads to  $\mu>1$.  


The stationary solution  for the magnetic field  which is current-free in the fluid is 
$
 B_\phi=B_{\rm in}R_{\rm in}/{R}$. We define
$
\mu_B={B_{\rm out}}/{B_{\rm in}},
$
 hence $\mu_B=\rin$.  For pinch-type solutions due to  homogeneous axial electric currents with $
 B_\phi=B_{\rm in}R/R_{\rm in}$  it is $\mu_B=1/\rin$.

The dimensionless physical parameters of the system besides the magnetic Prandtl number are  the 
Hartmann number $\Ha$ and the Reynolds number $\Rey$,
\begin{eqnarray}
 {\Ha} =\frac{B_{\rm in} D}{\sqrt{\mu_0\rho\nu\eta}},  \quad\quad\quad
 {\Rey} =\frac{\Om_{\rm in} D^2}{\nu}.
\label{pm}
\end{eqnarray}
The difference   $D=R_{\rm out}-R_{\rm in}$ 
is the gap width between the cylinders. The Hartmann number is defined with the magnetic field  at the inner wall. 
The ratio of the angular velocity of rotation and the \A\ frequency of the magnetic field is  the magnetic Mach number $\Mm$ which easily can expressed by the magnetic Prandtl number, the Reynolds number and the Hartmann number, i.e.
\begin{eqnarray}
 {\Mm} =\frac{\Om_{\rm in}}{\Om_{\rm A,in}} = \frac{\sqrt{\Pm}\Rey}{\Ha}.
\label{mm}
\end{eqnarray}
Fast rotation compared with the magnetic field is described by $\Mm>1$ and slow rotation by $\Mm<1$. $\Mm=1$ may be called  a magnetic sonic point. Many cosmical objects can be characterized by $\Mm>1$.

The variables  $\vec{U}$, $\vec{B}$ and $P$ are split into mean and fluctuating components, i.e. $\vec{U}=\bar{ \vec{U}}+\vec{u}$, $\vec{B}=\bar{ \vec{B}}+\vec{b}$ and $P=\bar P+p$. The bars from the mean-field variables are immediately dropped, so that the capital letters $\vec{U}$, $\vec{B}$ and $P$ represent the  background quantities. 
Simplifying, the nonaxisymmetric components of flow
and field may be used in the following as the ‘fluctuations’
while the axisymmetric components are considered as the
mean quantities. Then the averaging procedure is 
the integration over the azimuth $\phi$.
By developing the fluctuations  $\vec{u}$, $\vec{b}$ and $p$ into normal modes, 
$
[\vec{u},\vec{b},p]=[\vec{u}(R),\vec{b}(R),p(R)] {\rm exp}({{\rm i}(\omega t+kz+ m\phi)})$,
 the solutions of the linearized MHD equations are considered for axially unbounded cylinders. Here $k$ is the axial wave number of the perturbation, $m$ its azimuthal wave number and $\omega$ the complex frequency including growth rate as its negative imaginary part and  a drift  frequency $\omega_{\rm dr}$ 
 as its real part. 
A numerical  code is used to solve the resulting  set of linearized ordinary differential equations 
for the radial functions of flow, field and pressure  fluctuations. The solutions are
optimized with respect to the Reynolds number for given Hartmann number by varying the wave
number. Only solutions for $|m|=1$ are here discussed. For consistency only  such small Hartmann numbers are considered for which only these lowest unstable modes  are excited.
\subsection{Boundary conditions}
The hydrodynamic boundary
conditions at the cylinder walls are the rigid ones,  i.e. $u_R=u_\phi=u_z=0$. The cylinders are either considered as
perfectly conducting or insulating. For the conducting walls the
fluctuations  must fulfill ${\rm d} b_\phi/{\rm d}R + b_\phi/R=b_R=0$ at  $R_{\rm in}$
and $R_{\rm out}$ so that ten boundary conditions exist for the set of ten differential equations.
The magnetic boundary conditions for insulating walls are much more complicated, i.e.
\begin{equation}
b_R+\frac{{\rm i}b_z}{I_m(kR)} \left(\frac{m}{kR} I_m(kR)+I_{m+1}(kR)\right)=0
\label{72.7}
\end{equation}
for $R=R_{\rm in}$ and 
\begin{equation}
b_R+ \frac{{\rm i}b_z}{K_m(kR)} \left(\frac{m}{kR} K_m(kR)-K_{m+1}(kR)\right)=0
\label{72.8}
\end{equation}
for $R=R_{\rm out}$, where $I_m$ and $K_m$ are the modified Bessel functions of second kind. The conditions for the toroidal field  are simply $k R b_\phi =m\, b_z$ at $R_{\rm in}$ and $R_{\rm out}$. 
More details including the modified expressions for cylinders with {\em finite} electric conductivity have been given by \cite{RG18}.

\begin{figure}
 \centering
   \includegraphics[width=0.7\textwidth]{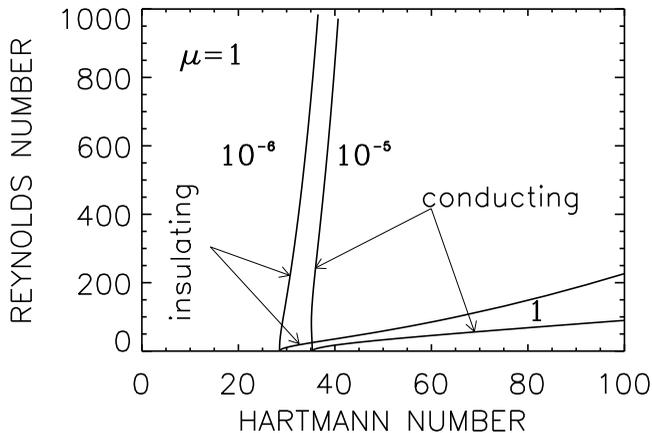}
 \caption{Stability map $\Remax=\Remax(\Ha)$ for  $z$-pinch with rigid rotation for small and large magnetic Prandtl numbers (marked). Models with parameters below the lines are unstable.  Both boundary conditions:   perfect-conducting  as well as  insulating   cylinders. $\mu=1$,  $\mu_B=2$, $\rin=0.5$,  $m=\pm 1$. } 
\label{fig1} 
\end{figure}

For the  magnetic field with  $\mu_B=1/\rin$ the  Fig.~\ref{fig1} shows the lines of neutral  stability (i.e. for vanishing growth rate) for the rigidly rotating  flow ($\mu=1$) for both sorts of boundary conditions.  The Tayler  instability for $m=\pm 1$  exists for  supercritical Hartmann numbers.
  Absolute   minima $\Ha_0$ of the
Hartmann numbers exist below which  the rotation law is stable. 
The curve of neutral instability  limits the instability  by suppressing the nonaxisymmetric field mode by too fast rotation. 

  To demonstrate  the influence of the boundary condition the  Fig. \ref{fig1}   gives
the instability lines   for  containers  with perfect-conducting  and insulating cylinders. The $\Ha_0=28.1$  for vacuum  conditions   prove to be smaller than the $\Ha_0=35.3$ for  perfect-conducting  conditions. With insulating cylinders the magnetized Taylor-Couette flows become more easily unstable than with conducting cylinders. 


As already noted   the magnetized  flow  of Fig. \ref{fig1} constitutes a standard  example of  the  Chandrasekhar-type flows.
The relations $\Om\propto R^{-q}$ and $B_\phi\propto R^{1-q}$ defining these  flows in Taylor-Couette systems  lead to the \C\ condition 
\beg
\mu=\rin\mu_B,
\label{chancon}
\ende 
hence  $\mu_B=2 \mu$  for $\rin=0.5$. This general condition  is fulfilled for the rigidly rotating $z$-pinch with $\mu=1$ and $\mu_B=2$. The lines of marginal stability of the flow basically coincide in the ($\Ha/\Rey$) plane for $\Pm\to 0$. All curves for small $\Pm$  of Fig. \ref{fig1} are thus identical (and invisible). 

The  magnetic Prandtl number of $\Pm=10^{-5}$ characterizes liquid sodium as the conducting fluid. The plotted lines of marginal instability  are valid for all  $\Pm\leq 10^{-5}$ including $\Pm=0$. They thus can  also be  obtained by use of  the inductionless approximation.

The $z$-pinch is characterized by  a homogeneous electric current in axial direction which  becomes unstable even without rotation if the current is strong enough.  
As is also demonstrated by the instability map 
the numerical value of  $\Ha_0$  does (slightly) depend on the boundary conditions but does not depend on the magnetic Prandtl number. 
On the other hand, however, the rotational suppression of the Tayler instability strongly depends on the magnetic Prandtl number.  For the very small $\Pm$ it almost disappears. The curves in Fig \ref{fig1} can also be characterized by  magnetic Mach numbers  which are almost independent of  the numerical value of the Hartmann number.  Obviously the magnetic Mach number slightly exceeds unity for $\Pm\simeq 1$ but it becomes smaller for smaller $\Pm$. As it is true for all \C-type flows (which scale with $\Ha$ and $\Rey$ for $\Pm\to 0$)  the magnetic Mach number decreases with decreasing $\sqrt{\Pm}$ for very small $\Pm$ which limits the astrophysical  relevance of the instability for objects of small magnetic Prandtl number.
\subsection{Angular momentum transport}
It is known that the radial angular momentum transport by  instability patterns can be described  by the ($R\phi$) component of the total stress  tensor,
\beg
T_{ij}= \langle u_i u_j\rangle - \frac{1}{\mu_0\rho} \langle b_ib_j\rangle+ \frac{1}{2\mu_0\rho}\langle \vec{b}^2\rangle \delta_{ij},
\label{energy}
\ende
as the difference
\beg
T_{R}= \langle u_R u_\phi\rangle - \frac{1}{\mu_0\rho} \langle b_R b_\phi\rangle
\label{rad}
\ende
 of the Reynolds and Maxwell stress tensors taken at the same spatial and temporal coordinates.  In the diffusion approximation (\ref{Bou}) this tensor component  is assumed as existing only for non-uniform rotation, representing  angular momentum as flowing in the direction of slower rotation (for positive $\nuT$). To probe the applicability of the diffusion approximation we shall compute  (\ref{rad}) for magnetic instabilities. 
As the  Tayler instability  even exists for rigid rotation its angular momentum transport should vanish. We shall calculate the radial flux of the angular momentum (\ref{rad}) along the lines of neutral  stability where it is allowed to use the linearized MHD equations.

\section{Rotating z-pinches}\label{Rotating}
The  $z$-pinch  is formed by an uniform electric current throughout the entire region $R<R_{\rm out}$. Any resulting instability is purely current-driven, it even exists  for $\Rey=0$, the rotation only acts suppressing  \citep{PT85}.
The curves of Fig. \ref{fig1}  demonstrate the stabilizing effect of  rotation    which is strongest for $\Pm=1$. It becomes weaker for smaller magnetic Prandtl numbers.
 In all cases a maximal Reynolds number $\Remax$ exists for given Hartmann number  above which the $z$-pinch is stable. The  $\Remax$ defining the Reynolds number of neutral stability depends on the magnetic Prandtl number, i.e. the smaller the $\Pm$ the higher the $\Remax$.  
  Quasi-Keplerian flow of  $ \Ha=50$ allows instability only up to  $\Remax\simeq 1000$ for $\Pm=1$ and $\simeq 3000$ for $\Pm=0.01$ independent of the used  boundary conditions.

\subsection{The eigenfunctions}
\begin{figure}
 \centering
  \hbox{ 
 \includegraphics[width=0.33\textwidth]{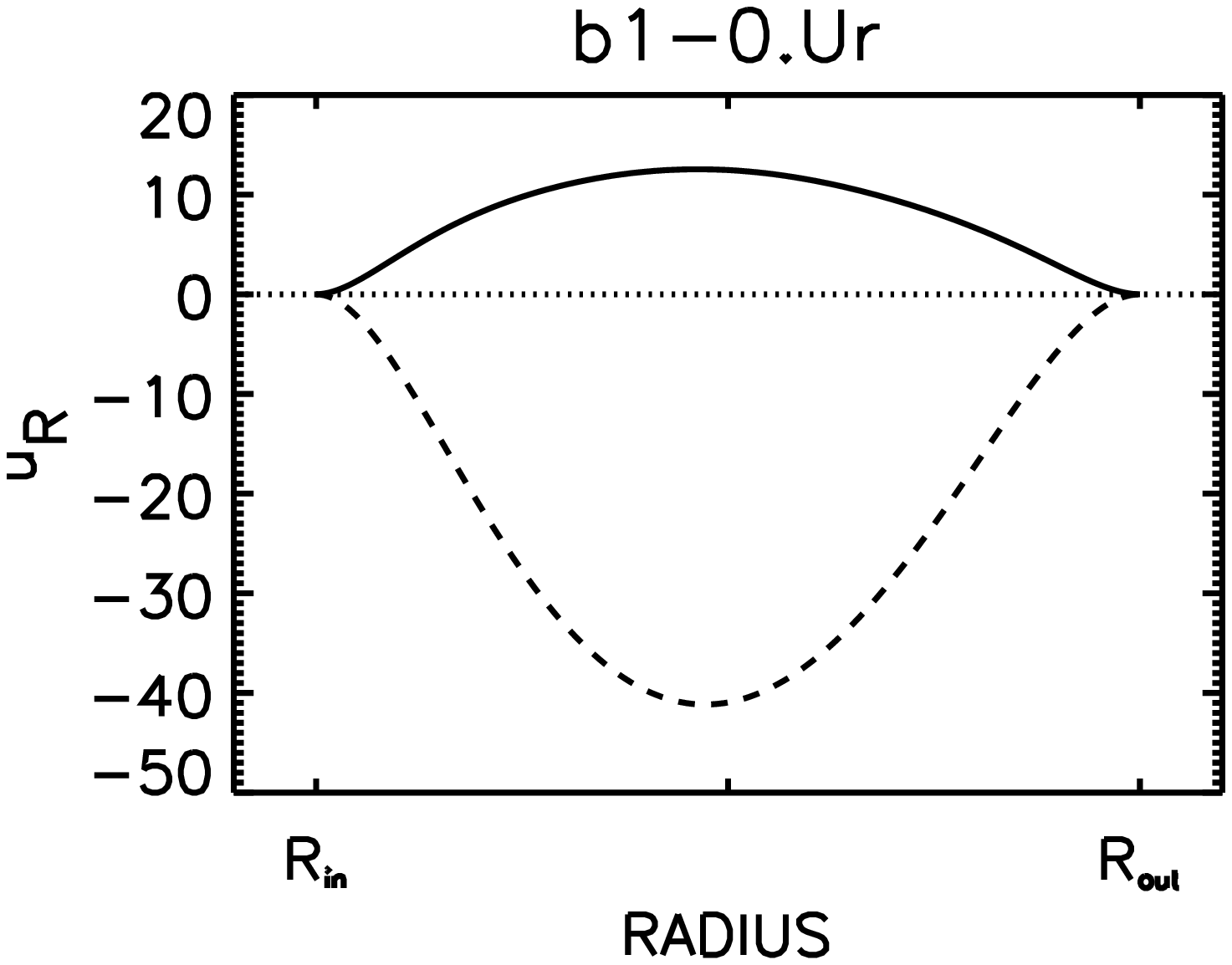} 
 \includegraphics[width=0.33\textwidth]{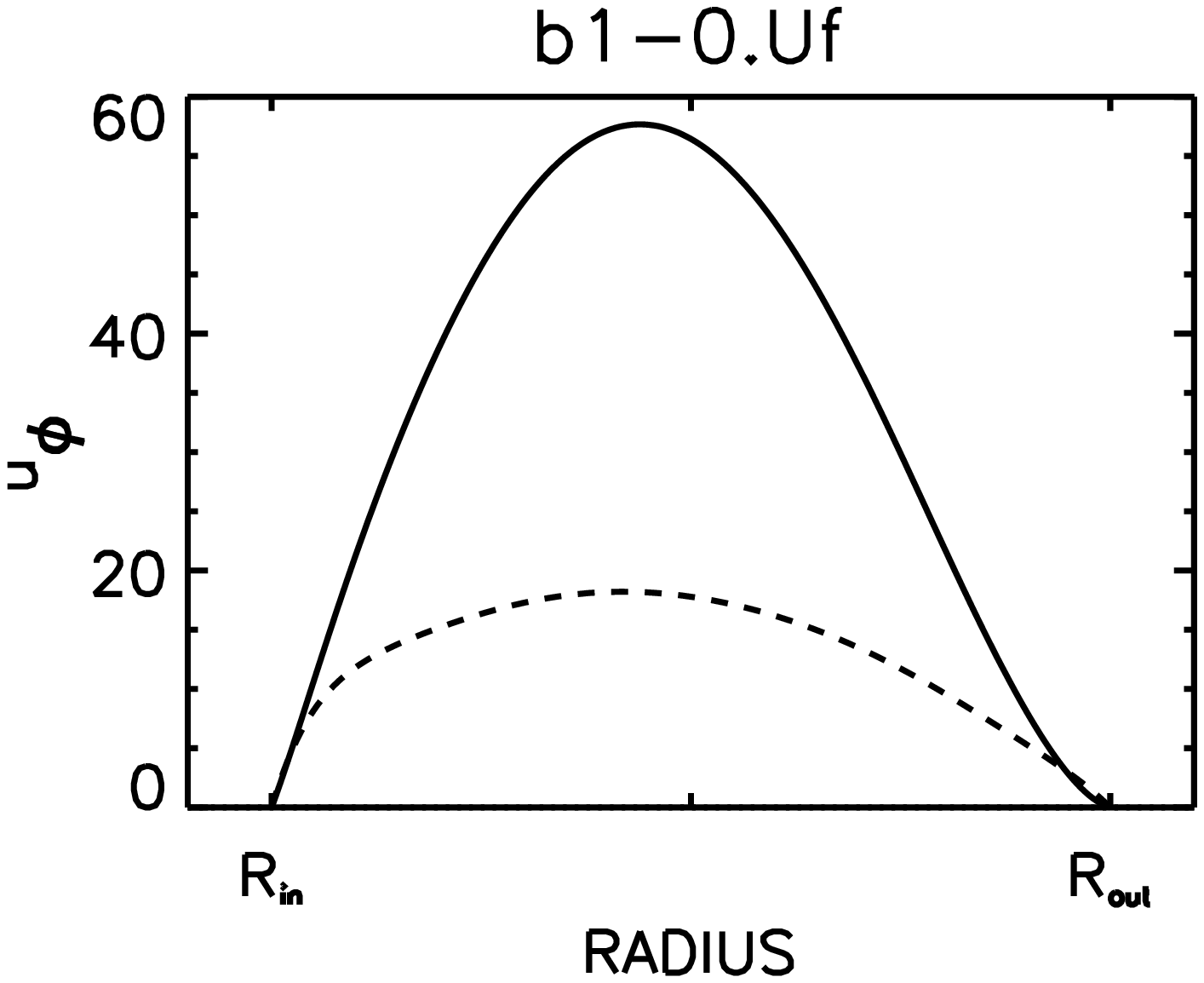} 
 \includegraphics[width=0.33\textwidth]{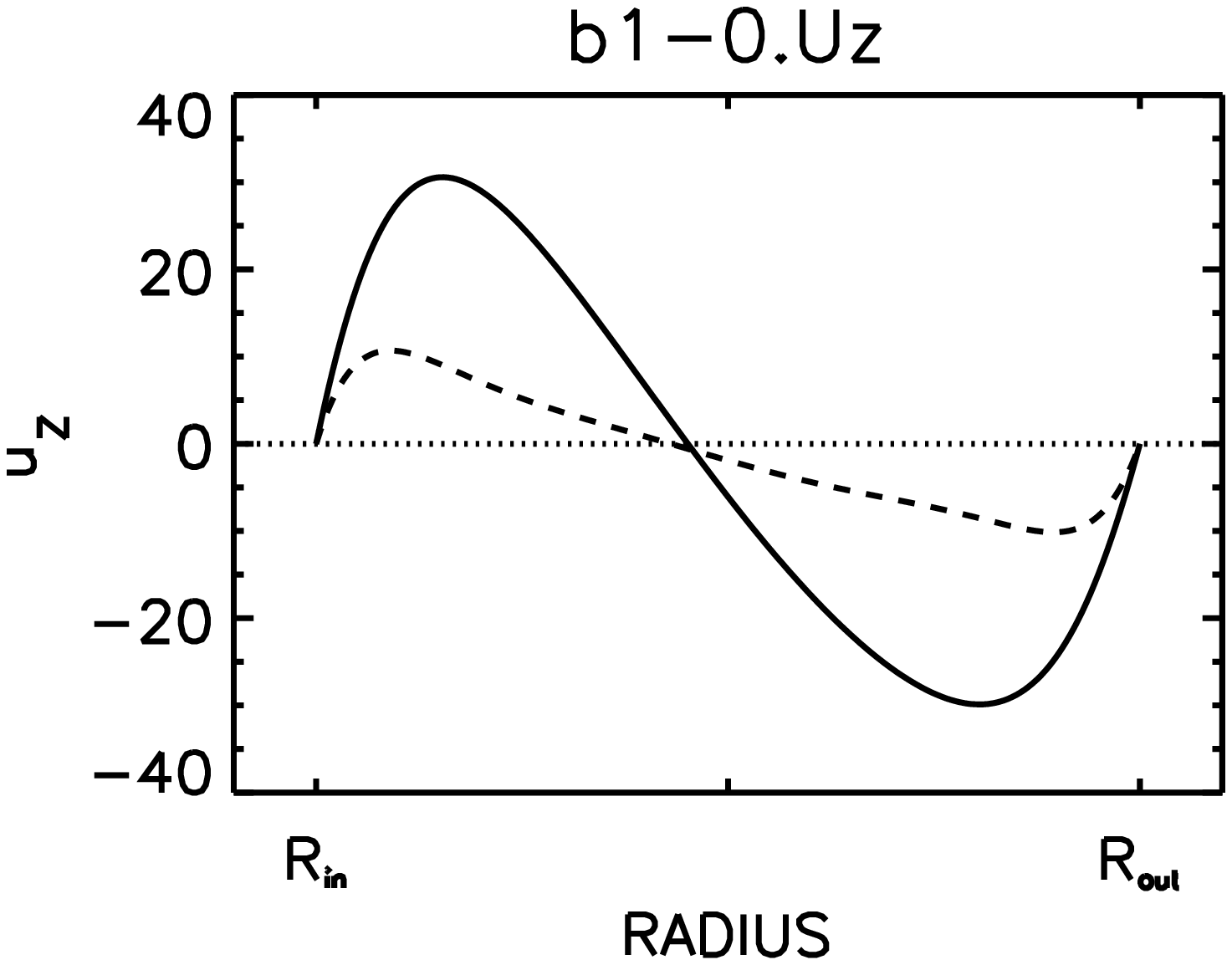} 
 }
 \hbox{ 
 \includegraphics[width=0.33\textwidth]{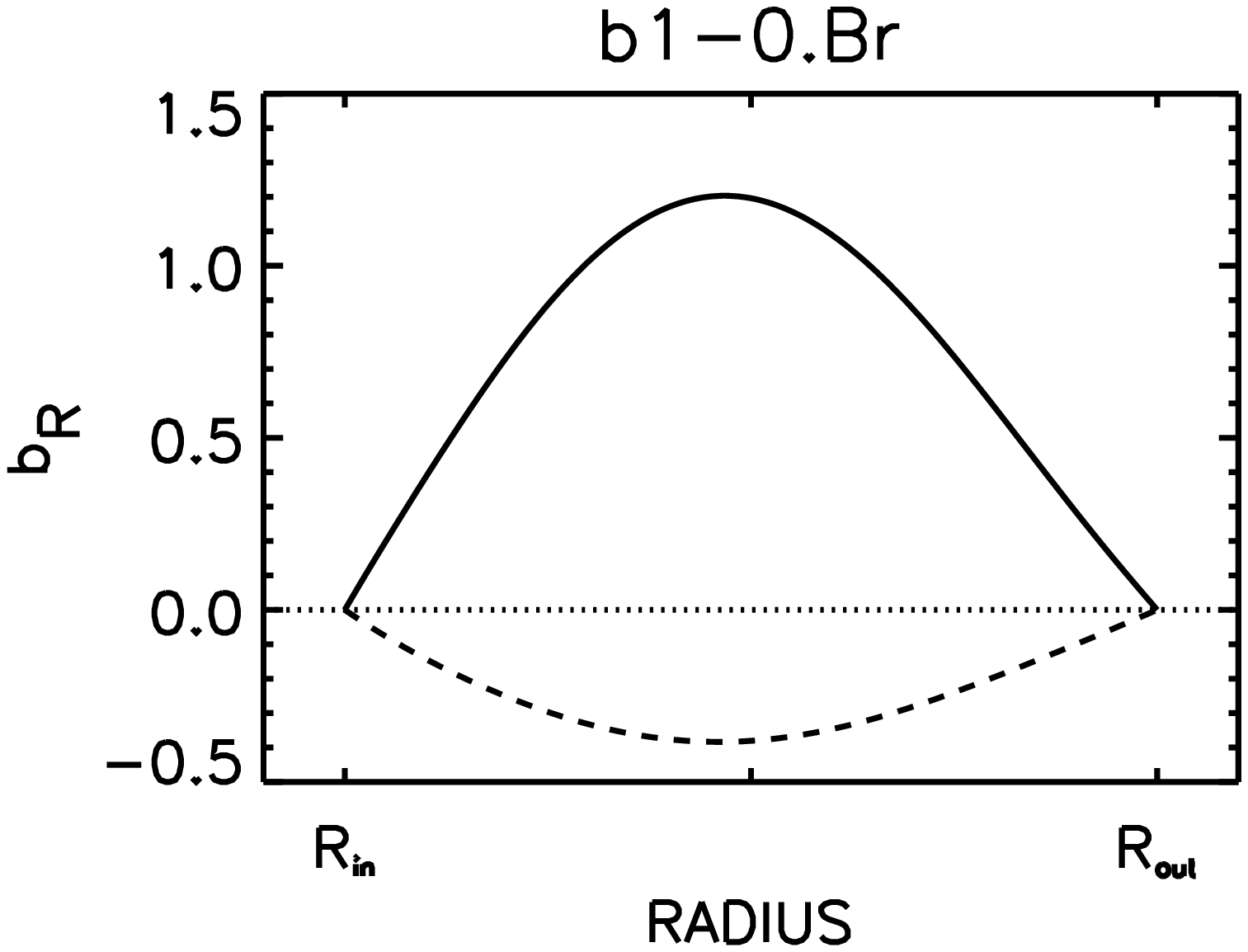} 
 \includegraphics[width=0.33\textwidth]{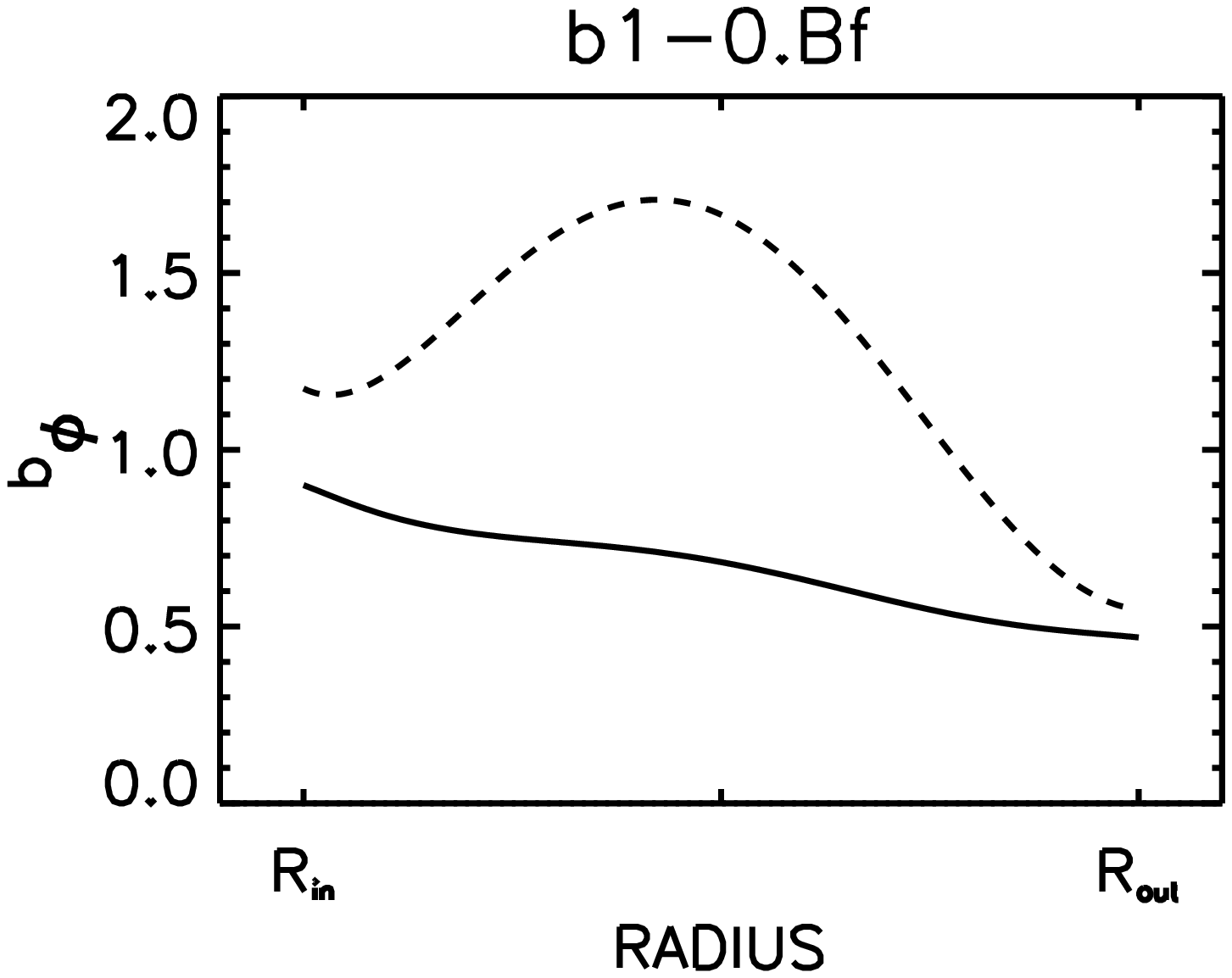} 
 \includegraphics[width=0.33\textwidth]{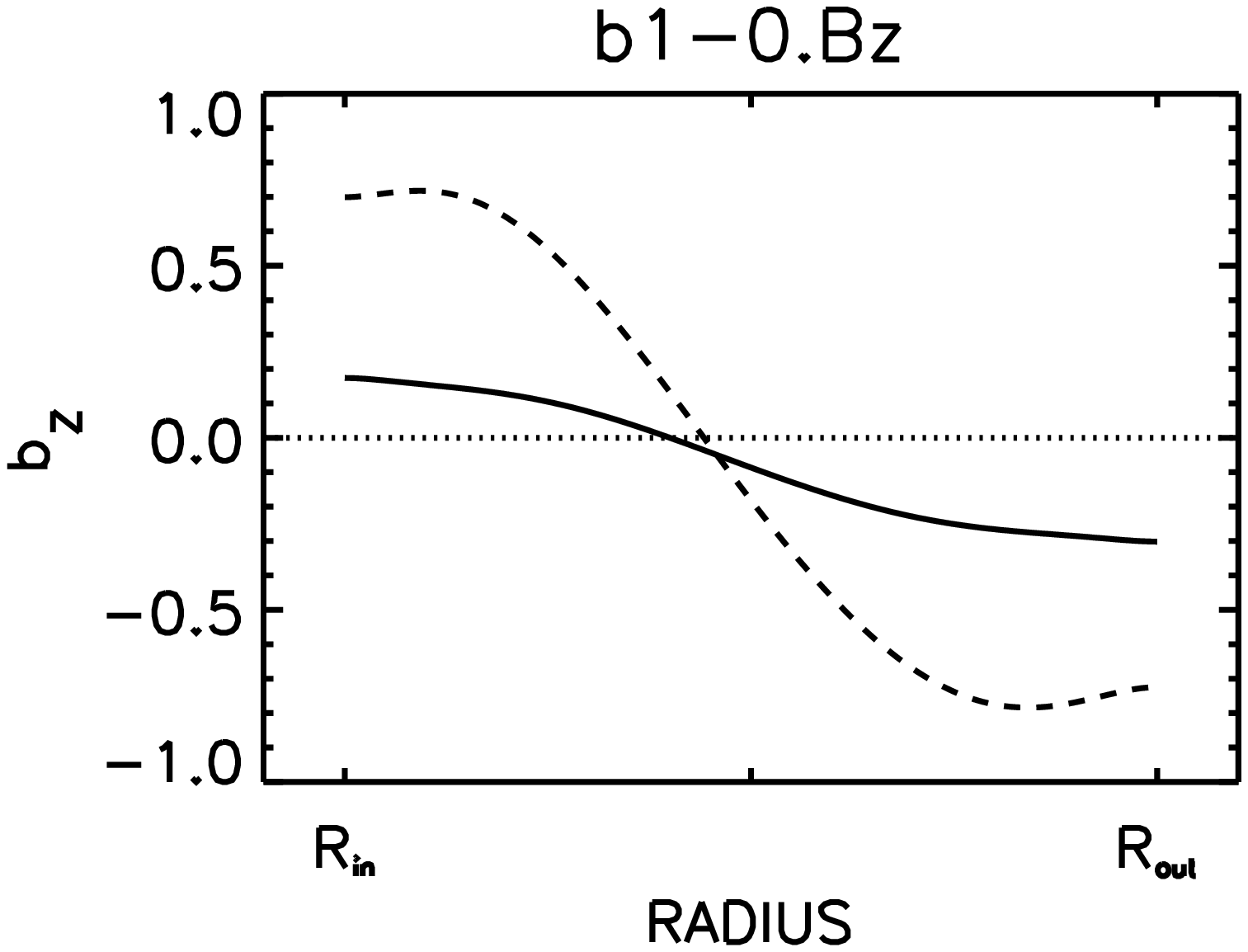} 
 }
 \caption{Eigenfunctions (real parts: solid lines; imaginary parts: dashed lines) of the rigidly rotating $z$-pinch for $\Ha=50$ and $\Rey=144$  for  perfect-conducting boundary conditions.  $\rin=0.5$. $m=  1$, $\Pm=0.1$, $\mu_B=2$, $\mu=1$. } 
\label{fig3} 
\end{figure}

The homogeneous system of differential equations for the perturbations forms  an eigenvalue problem with eigensolutions for $\vec{u}(R)$ and $\vec{b}(R)$ which can be determined up to  a free real multiplication factor. The sign of products of two  perturbation components, therefore,  remains unchanged. For the mode with $m=1$ these functions in their dependence on $R$ are given in Fig. \ref{fig3} for $\Pm=0.1$. Note  the boundary conditions $\vec{u}=0$ and $b_R=0$ (for perfect-conducting  walls) as fulfilled.
One also finds, as it must,  for $m=-1$ the components $u_R, u_\phi, b_R$ and $b_\phi$ are conjugate-complex as also  the field components $- {\rm i} u_z$ and $- {\rm i} b_z$ are. It means that for the transformation $m\to -m$ the components $b_R$  transform as $b_R^{\rm R}\to b_R^{\rm R} $ and $b_R^{\rm I}\to -b_R^{\rm I} $ (the same for $b_\phi$) while for  $m\to -m$ it is  $b_z^{\rm R}\to - b_z^{\rm R} $ and $b_z^{\rm I}\to b_z^{\rm I} $.  The superscripts R stand for the real parts and I for the imaginary parts of the eigensolutions. 

The product of two scalars $A$ and $B$ after averaging over the $\phi$ coordinate is the sum of the products of the real parts and the imaginary parts, i.e. $AB = A^{\rm R} B^{\rm R}+  A^{\rm I} B^{\rm I}$. There is a certain factor in front of this expression whose value, however,   is unimportant as in the linear theory the  vector components are only known   up to a free factor.  In Fig. \ref{fig3}  the magnetic-induced  contribution $b_R b_\phi$ to the radial flux of angular momentum is positive.  One expects  $b_R b_\phi\simeq 0$ for rigid rotation which, however,  is here not the case.
As the kinetic  transport $ u_R u_\phi$  is (slightly) negative the total stress component $T_R$ is  negative, too.

\begin{figure}
  \centerline{ 
    \includegraphics[width=0.5\textwidth]{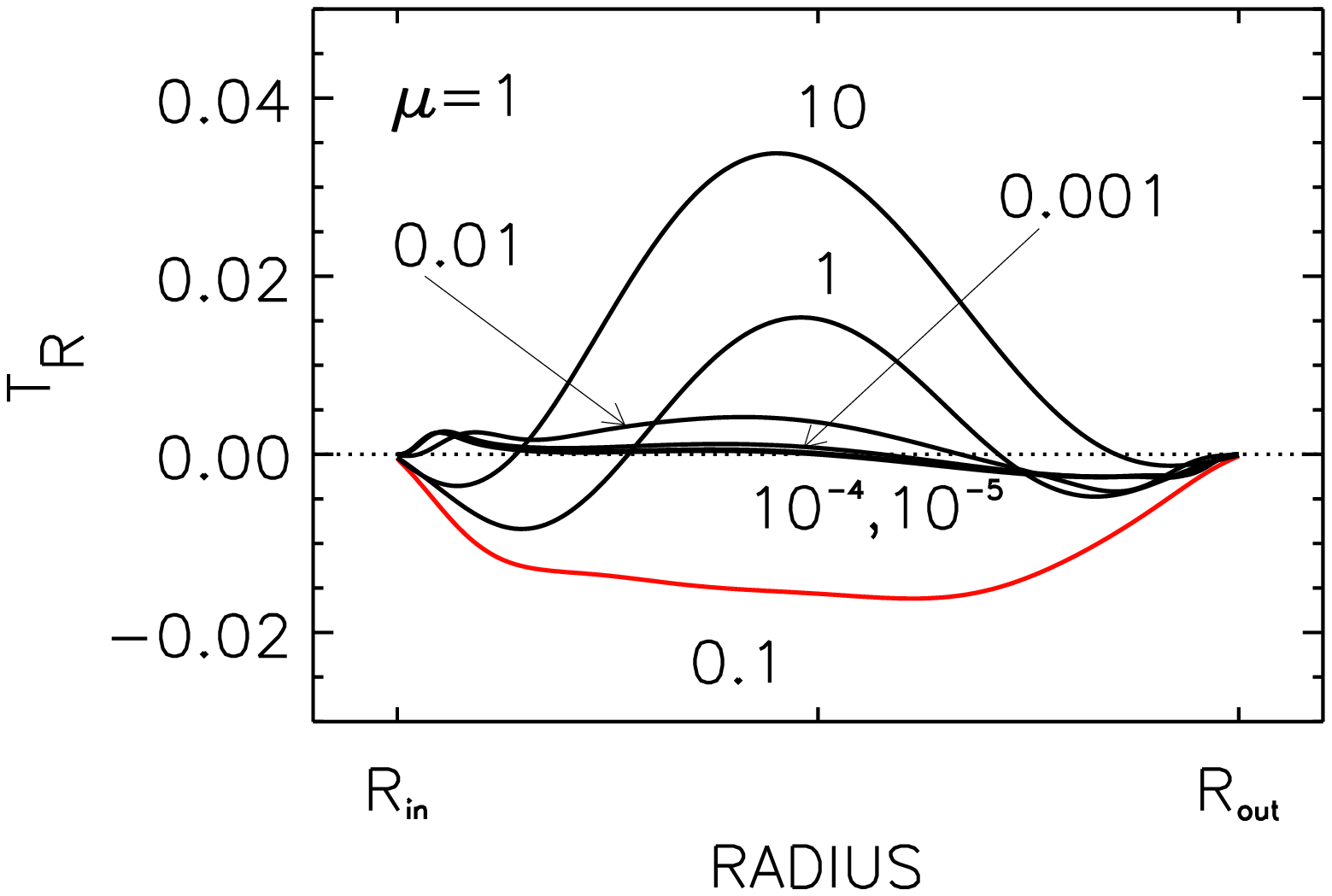}
      \includegraphics[width=0.5\textwidth]{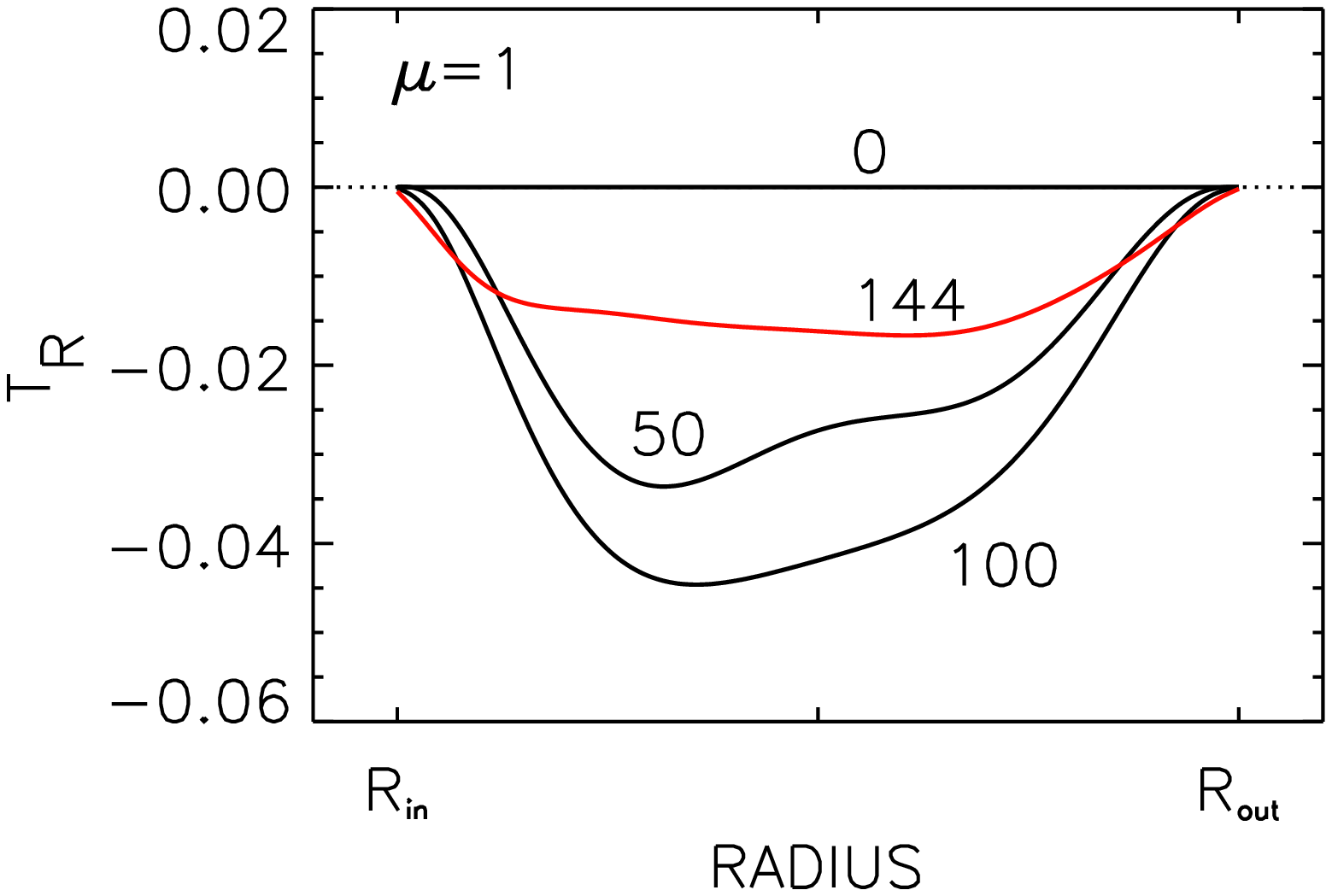}
 }
  \caption{Normalized angular momentum transport $T_R(R)$ as a function of the radius $R$ for the $z$-pinch with rigid rotation. Left: variation of $\Pm$ (marked) for neutral stability ($\Rey=\Remax$), right: variation of $\Rey$ (marked) for fixed $\Pm=0.1$ (vertical cut in the stability map). The curves are identical for $m\to -m$.   $\mu=1$, $\rin=0.5$, $\mu_B=2$, $m= 1$,   $\Ha=50$ (always).  Perfect-conducting boundary conditions.}
\label{fig32}
\end{figure}

The plotted amplitudes of the functions are completely arbitrary, they have  no physical meaning. The above given sign rules for the products of radial and azimuthal components of flow and fields  are not influenced by the boundary conditions. 
As with our normalizations the total stress results from the difference $\langle u_R u_\phi\rangle- {\Pm\Ha^2} \langle b_R b_\phi\rangle$ (where  the instability sets in for almost constant $\Ha$ if $\Pm\to 0$) the sign of the angular momentum transport is determined by the flow perturbations for small $\Pm$ and by the magnetic fluctuations for large $\Pm$.
The torque on the cylinders obviously vanishes if by the boundary conditions $u_R=b_R=0$. The  average procedure may concern the cylinder surface formed by $z$ and $\phi$.  Then it becomes   $\langle u_R u_\phi\rangle \propto u^{\rm R}_R u^{\rm R}_\phi + u^{\rm I}_R u^{\rm I}_\phi$. All the terms mixed in R and I disappear after averaging over the azimuth $\phi$. The same procedure holds for the magnetic terms.
From the   transformation rules for $m\to -m$ it is clear that the  modes $m$ and $-m$ lead to the same flux of angular momentum. The curves for $T_R$  in Fig. \ref{fig32} are thus  identical for $m=1$ and $m=-1$.

We have  to normalize the expression (\ref{rad}) in order to compensate the role of   the free parameter in the eigenfunctions. To this end the $T_R$ is divided by the total energy 
$E=\langle\vec{u}^2\rangle+\langle\vec{b}^2\rangle /\mu_0\rho$
(in our units  $E=\langle\vec{u}^2\rangle+ \Pm \Ha^2 \langle\vec{b}^2\rangle $).
According to this definition the normalized $|T_{R}|$ should not exceed unity.

 The angular momentum transport by
 the current-driven Tayler instability with $\mu_B=2$ and  $\mu=1$ is given by Fig. \ref{fig32}  
 for a fixed Hartmann number; $\Ha=50$. The  flow is of the Chandrasekhar-type and because of its uniform rotation    angular momentum should not be transported but it does. Note the negative value for  $\Pm\lsim 0.1$ which has been suggested  above by inspection of Fig. \ref{fig3}. Generally, the transport direction   depends on the magnetic Prandtl number  but its absolute value  $|T_R|$ becomes very small for the smallest magnetic Prandtl numbers.
After Eq. (\ref{Bou}) the angular momentum transport $T_R$ should vanish for all values of the parameters if the $z$-pinch rotates with uniform $\Om$ but this  is not confirmed by the calculations.
\begin{figure}
  \centerline{ 
   \includegraphics[width=0.6\textwidth]{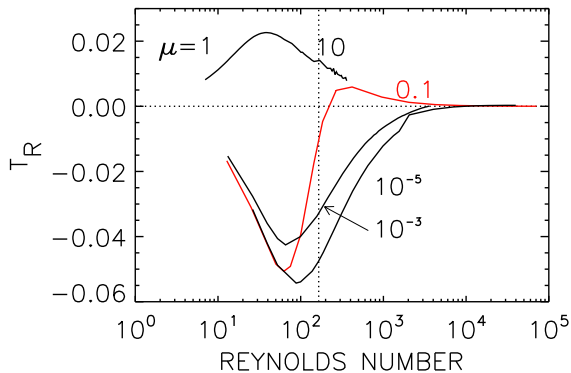}
 }
  \caption{Normalized angular momentum transport $T_R$ (averages in the entire container) for the rigidly rotating $z$-pinch along the line of neutral instability $\Rey=\Remax$.  The curves are marked with  $\Pm$.  
 The radial momentum flux   vanishes close to    the magnetic sonic point  (the vertical dotted line) for $\Pm=0.1$. The $T_R$ curves are invariant against the transformation   $m\to -m$. $\mu=1$, $\mu_B=2$,  $\rin=0.5$, $m=  \pm 1$. Perfect-conducting boundary conditions.}
\label{figamt}
\end{figure}

\subsection{$\Lambda$ effect for rigid-body rotation}\label{Rigid}
The left panel of Fig. \ref{fig32} gives  the normalized radial fluxes of angular momentum between the two cylinders for the rigidly rotating $z$-pinch for various magnetic Prandtl numbers and for one and the same (small)  Hartmann number, $\Ha=50$\footnote{the maximal Reynolds number  for $\Pm=0.1$ is $\Remax=144$}. 
It is positive for   large $\Pm$,  it is negative for   $\Pm\lsim 0.1$  and it almost vanishes for  very small $\Pm$. As we know,  for very small $\Pm$  the angular momentum transport is dominated by the Reynolds stress. 
while for large $\Pm$  the magnetic terms exceed the kinetic ones. 
Then only the   negativity of the Reynolds stress produces  negative $T_R$. For large magnetic Prandtl numbers it can  be overcompensated by negative Maxwell stress $\langle b_R b_\phi\rangle$. 
Note, that there is no shear which could explain the anticorrelation of the  fluctuations $b_R$ and $b_\phi$.

The right panel of Fig. \ref{fig32} demonstrates that the calculations along the line of neutral stability ($\Rey=\Remax$) may underestimate the numerical values of the possible (normalized) angular momentum fluxes. The $T_R(R)$ is computed along a vertical cut for $\Ha=50$ and $\Pm=0.1$. The curves are marked with the used values of the Reynolds number. The two red lines in both panels of Fig.\ref{fig32} are identical. Of course, $T_R=0$ for $\Rey=0$. The remaining curves for $\Rey= 50$ and $\Rey=100$ yield larger numerical values as the curve for the neutral line, for $\Rey=\Remax$. The growth rates of the instability pattern for these curves  do not vanish.

Figure \ref{figamt} presents the normalized angular momentum transport $T_R$ for rigid rotation along the lines of neutral stability for several values of the magnetic Prandtl number. The $T_R$ is now averaged over the entire container.  The vertical dotted line gives for $\Pm=0.1$ the magnetic sonic point where $\Mm=1$. For the same $\Pm$ the red line shows positive values for large magnetic Mach numbers and negative values for small magnetic Mach numbers. The figure    also provides  nonvanishing  $T_R$ for $\Pm\neq 0.1$.  For $\Pm<0.1$ even for faster rotation the  $T_R$ remains negative. For  $\Pm=10$ the $T_R$ is also finite but it is positive. The  reason is that the magnetic sonic point for $\Pm=10$ occurs already  for $\Rey=13$ hence the main part of the corresponding curve belongs to super-\A ic flows. For $\Pm=10^{-6}$ the magnetic sonic point is located at $\Rey=\simeq 20.000$.

\begin{figure}
  \centerline{ 
    \includegraphics[width=0.55\textwidth]{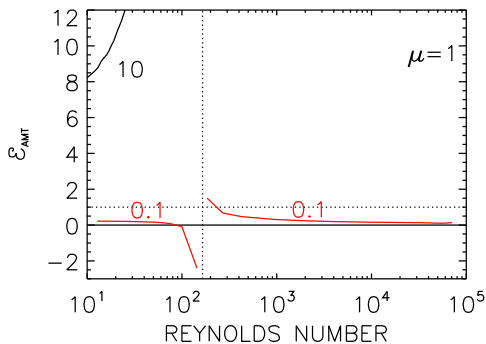}
 }
  \caption{Stress ratios  (\ref{eps}) along the lines of neutral stability 
  for $\Pm=0.1$ and $\Pm=10$ with averages over the container. The horizontal dotted line gives $\varepsilon_{\rm AMT}=1$,  the vertical  line denotes   the magnetic sonic point ($\Rey= 180$) for $\Pm=0.1$.  $\rin=0.5$, $\mu=1$, $\mu_B=2$, $m=  1$. Perfect-conducting boundary conditions.}
\label{figamt1}
\end{figure}

For a more detailed discussion  Fig. \ref{figamt1} gives the ratio
 \beg
\varepsilon_{\rm AMT}=\frac{\langle b_R b_\phi\rangle}{\mu_0\rho \langle u_R u_\phi\rangle} 
 \label{eps}
 \ende
   of the Maxwell stress and the Reynolds stress versus the Reynolds  number for the containers  with $\Pm=0.1$ and $\Pm=10$. Striking differences only  exist  in the vicinity of the magnetic sonic point with $\Mm=1$ (vertical dotted lines in Figs. \ref{figamt} and \ref{figamt1}). For larger as well as for smaller Reynolds  numbers it is  $0<\varepsilon_{\rm AMT}<1$, hence the Reynolds stress (slightly) exceeds the Maxwell stress. For the (normalized) stresses it is 
 \beg
 \langle u_R u_\phi\rangle \propto \frac{T_R}{1-\varepsilon_{\rm AMT}}, \ \ \ \ \ \ \ \ \ \ \ \ \ \ \ \  \frac{\langle b_R b_\phi\rangle}{\mu_0\rho}  \propto \varepsilon_{\rm AMT}\frac{T_R}{1-\varepsilon_{\rm AMT}}.
 \label{epseps}
 \ende

 For small $\varepsilon_{\rm AMT}>0$ it is $ \langle u_R u_\phi\rangle\propto T_R$ and also $\langle b_R b_\phi\rangle\propto  T_R$. Both stresses possess the same sign. This case is realized along almost the whole red line  of Fig. \ref{figamt1}. Left from the vertical dotted line both stresses are negative and right from the dotted line both stresses are positive.
 A  negative Maxwell stress for all sub-rotation laws is  easy to understand by the induction process due to differential rotation but it even exists for slow and rigid rotation.
On the other hand,  for large $|\varepsilon_{\rm AMT}|$ the Maxwell stress dominates  and it is simply
 $\langle b_R b_\phi\rangle\simeq -T_R$ so that $\langle b_R b_\phi\rangle>0$ for  $T_R<0$ (left from the dotted line) and  $\langle b_R b_\phi\rangle<0$ for  $T_R>0$.  The sign of the Maxwell stress changes where the red line crosses the horizontal solid line for $\varepsilon_{\rm AMT}=0$. 
 This is 

The zeros of $T_R$ in the   Fig. \ref{figamt} (horizontal dotted line) are defined by   $\varepsilon_{\rm AMT}=1$ which happens close to the sonic point  $\Mm= 1$.  For  larger $\Mm$ the  Reynolds  stress exceeds the Maxwell stress and it  is positive  leading to $T_R>0$.
 For  $\Mm\simeq 1$  the magnetic stress becomes  important and it is  $ \langle b_R b_\phi\rangle>0$ leading to negative $T_R$. For even smaller Mach numbers again the Reynolds stress exceeds the Maxwell stress but both are  now negative. 

Large values  $\varepsilon_{\rm AMT}>1$ for $\Pm=10$ shown in Fig. \ref{figamt1} indicate the dominance of the Maxwell stress for the  angular momentum transport. The positive $T_R$ values for large $\Pm$ result from a large anticorrelation of the magnetic perturbations $b_R$ and $b_\phi$ despite of the  rigid rotation.

  All these findings are rather general as  free parameters do not enter the calculations of   the ratio $\varepsilon_{\rm AMT}$. 
The question arises whether the diffusion approximation  (\ref{Bou})  must be modified by an additional  term 
\beg
T_{ij}=....+\Lambda_{ijk}\Om_k,
\label{Lambda}
\ende
in the stress tensor $T_{ij}$ which does not vanish for rigid 
rotation (``$\Lambda$ effect''). 
We have seen  that   indeed for the rigidly rotating $z$-pinch such a $\Lambda$ effect  exists  with a strong dependence on the magnetic Prandtl number.

The  $\Lambda$ effect is   known to appear  in rotating convective  spheres which are basically anisotropic in the radial direction $\vec g$ due to the density stratification. The tensor $\Lambda_{ijk}= (\epsilon_{ikl} g_j+\epsilon_{jlk}g_i)g_l$ appears in the stress tensor  leading to cross correlations $T_{r\phi}$ also at the equator. 
For unstratified but magnetized turbulences a very similar tensor may  exist after the transformation $\vec g\to \vec B$, i.e. $\Lambda_{ijk}= (\epsilon_{ikl} B_j+\epsilon_{jlk}B_i)B_l$. The radial flux of angular momentum for toroidal fields is thus $T_{R\phi}\propto B_\phi^2\Om$  suggesting that the results obtained  from   Fig. \ref{fig32} perform a new magnetic-induced realization of the $\Lambda$ effect.

The amplitudes of the normalized $T_R$ in Fig. \ref{fig32} for rigid rotation are smaller by one order of magnitude than those for quasi-Keplerian rotation (see below) but they are not very small. About 2\%
of the kinetic and magnetic energy are included in the cross correlation of radial and azimuthal components which, also in comparison with rotating convection,  does not seem as unreasonable. Depending  on the rotation rate \cite{K19} obtains cross correlations $Q_{r\phi}$ of 1...10 \% of the turbulence energy.
The sign of the perturbation-induced radial  fluxes $T_{R}$, however,  strongly depends on the value of $\Pm$ while for  rotating convection it is negative-semidefinite as numerical simulations show \citep{C01,HK06}. At the equator and for fast rotation, however,  the $T_{r\phi}$ generated by convection almost vanishes.

\section{Nonuniform rotation}\label{Nonuniform}
 Among the possible rotation laws the quasi-Keplerian rotation  is of particular interest in astrophysics. In a Taylor-Couette setup this flow is approximated by the assumption that the two cylinders rotate around the central  axis like planets, hence $\mu= \rin^{1.5}$. For $\rin=0.5$ the rotation ratio for such quasi-Keplerian flows  is $\mu=0.35$. 
This rotation may  be influenced by two different magnetic field profiles. 
The first  example is given by the $z$-pinch with  $\mu_B=2$  while the second model may fulfill the condition (\ref{chancon}), i.e. $\mu_B=0.7$ for $\rin=0.5$.
Only the latter model  belongs to the class of Chandrasekhar-type flows.

\begin{figure}
  \centerline{
 \vbox{
\hbox {\includegraphics[width=0.49\textwidth]{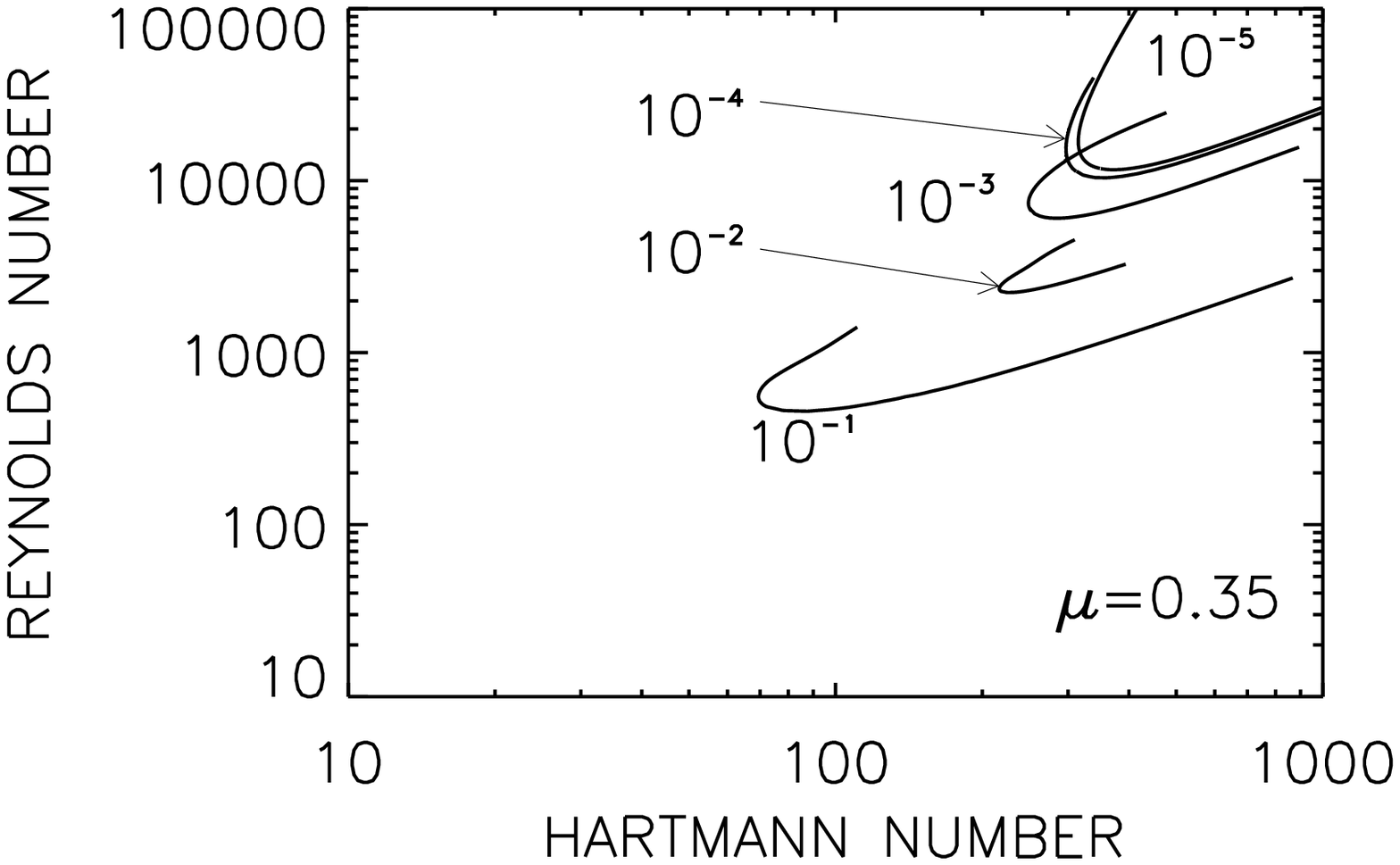}
   \includegraphics[width=0.49\textwidth]{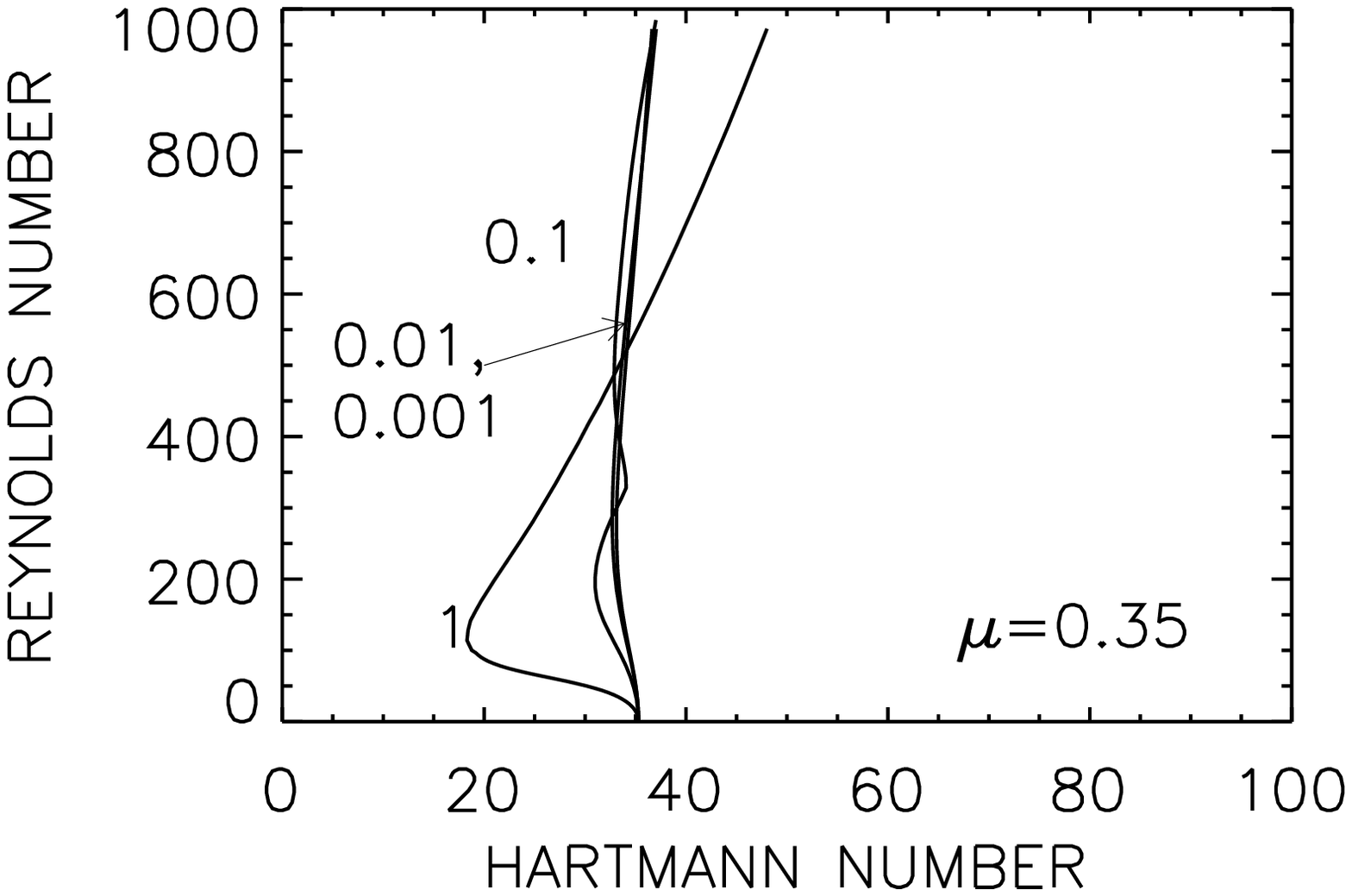}}
   \hbox{
    \includegraphics[width=0.49\textwidth]{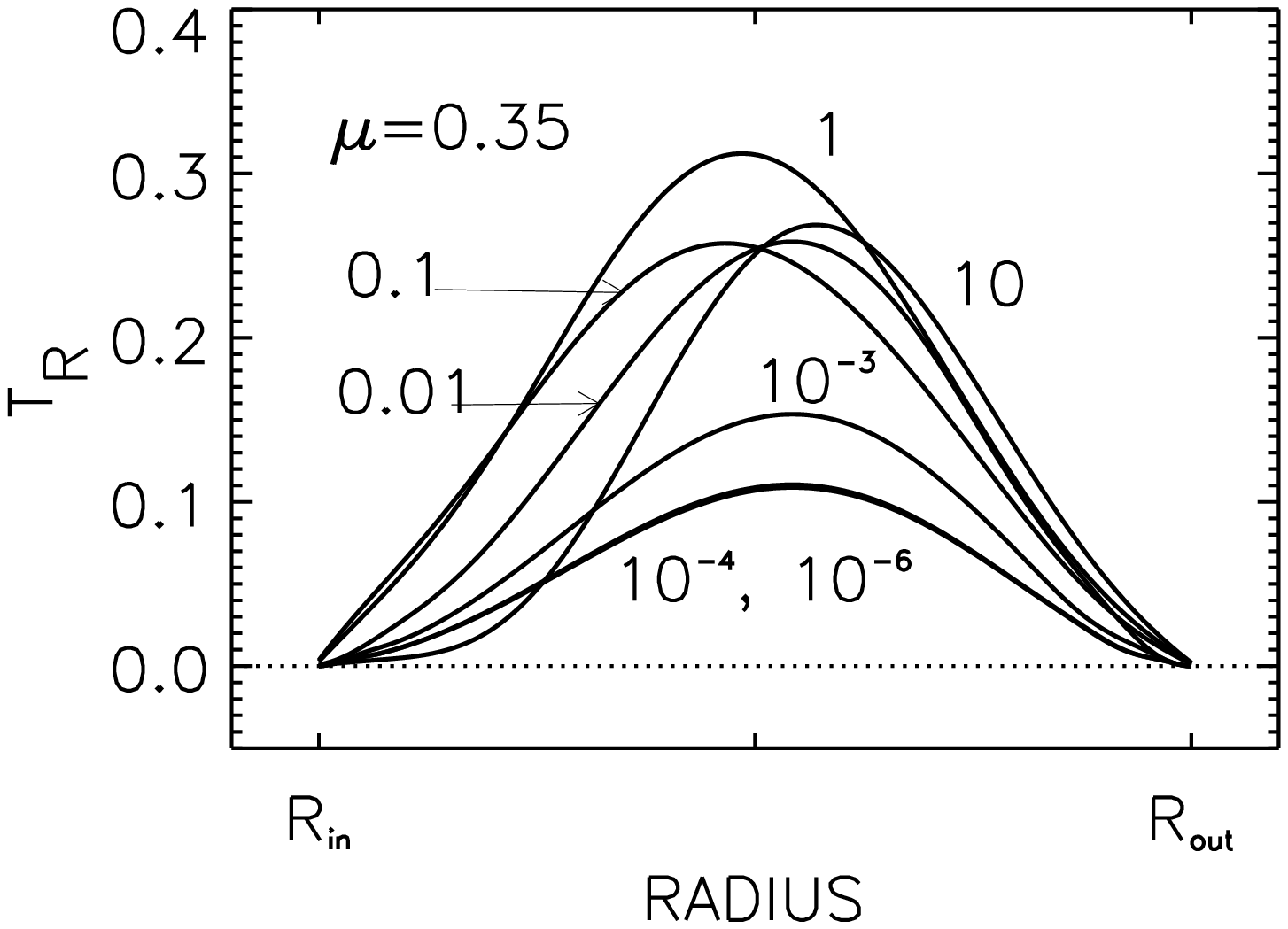}
   \includegraphics[width=0.49\textwidth]{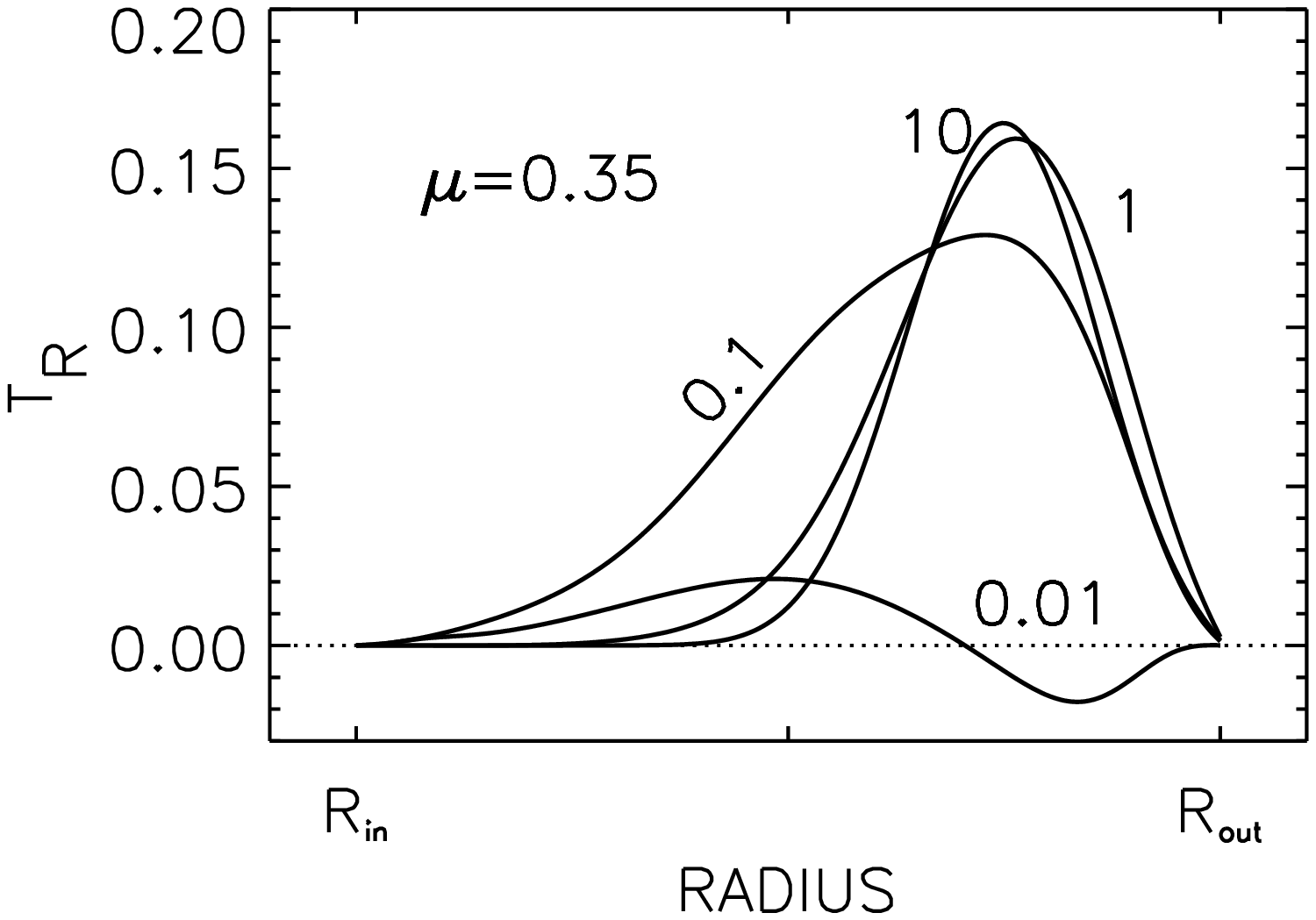}   }
   }  }
  \caption{Top: Stability maps for the quasi-Keplerian rotation  with   $\mu_B=0.7$ (left, \C-type) and $\mu_B=2$ ($z$-pinch, right). For the pinch it is $\Ha_0 = 35.3$. The curves are marked with their values of $\Pm$. 
  Bottom: Radial angular momentum transport by quasi-Keplerian rotation  for $\mu_B=0.7$ (left) and  $\mu_B=2$ (right).   The Hartmann numbers are  $\Ha=\Hamin$ for $\mu_B=0.7$ and $\Ha=50$ for $\mu_B=2$.
    $\rin=0.5$,  $\mu=0.35$, $m= \pm 1$. Perfect-conducting cylinders.}
\label{figkep}
\end{figure}

The stability map for the  field profiles fulfilling the \C\ condition combines typical properties of the maps for AMRI and Tayler instability.  The left top panel of Fig. \ref{figkep} gives  the lines of neutral stability  for the field with $\mu_B=0.7$. The curves for all $\Pm$ possess crossing points $\Ha_0$ with the horizontal axis ($\Rey=0$, not visible) but they also possess the absolute minimum  Hartmann numbers  $\Hamin$ with $\Hamin<\Ha_0$.  The curves for very  small $\Pm$ coincide. For small magnetic Prandtl numbers the $\Hamin$ (and also the associated Reynolds numbers) for the \C\ flow are  rather small. 
\C-type fields are thus  more unstable than fields which are current-free between the cylinders \citep{KS13}. 
Even more unstable is the $z$-pinch with $\mu_B=2$ combined with quasi-Keplerian rotation where the  instability  is  excited for very small Hartmann numbers $\Ha_0= 35.3$  (independent of the magnetic Prandtl number, top right  panel of  Fig. \ref{figkep}).

The  first question   is whether the quasi-Keplerian flow  with $\mu_B=0.7$ also provides an anomalous   angular momentum transport  as the rigidly rotating $z$-pinch which is also   of the Chandrasekhar-type.
For many $\Pm$ the angular momentum transport  has thus been calculated at  $\Ha=\Hamin$ (Fig. \ref{figkep}, bottom panel left) and for $\Ha=50$ (Fig. \ref{figkep}, bottom panel right) .  The results  provide positive $T_R$ for all  $\Pm$ without any  exception. An anomalous angular momentum transport does  thus not exist  for the two given examples with quasi-Kepler rotation law. 

\begin{figure}
  \centerline{
 \includegraphics[width=0.55\textwidth]{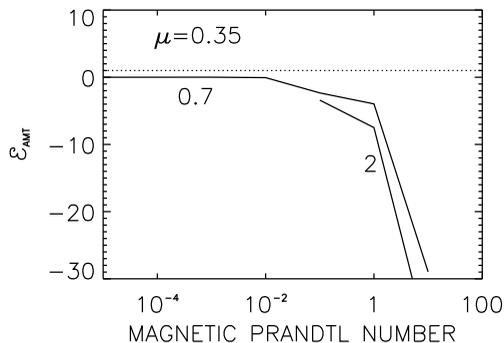}
  }
   \caption{
  Container-averaged stress ratios  (\ref{eps})  for the quasi-Keplerian flows  characterized  in Fig. \ref{figkep}.  The Hartmann numbers are  $\Ha=\Hamin$ for the \C-type flows ($\mu_B=0.7$) and $\Ha=50$ for the $z$-pinch ($\mu_B=2$).  The horizontal dotted line  symbolizes the ratio being unity,  $\varepsilon_{\rm AMT}=1$. 
  $\mu=0.35$,  $m=\pm 1$, $\rin=0.5$. Insulating cylinders.}
\label{fig44}
\end{figure}

In order to demonstrate  the basic  influence of the magnetic Prandtl number on the contribution  of the Maxwell stress to the angular momentum transport Fig.  \ref{fig44} gives the stress ratio  $\varepsilon_{\rm AMT}$  for the \C-type flow with quasi-Keplerian rotation and the $z$-pinch. Up to a magnetic Prandtl number of $10^{-2}$ the Maxwell stress is not important but  for larger $\Pm$ it is (because of $|\varepsilon_{\rm AMT}|\gg1$). Yet  for these examples  with growing magnetic Prandtl number also the magnetic Reynolds number grows. Moreover, one  always finds $\varepsilon_{\rm AMT}<0$ so that  $\langle u_R u_\phi\rangle>0$ and  $\langle b_R b_\phi\rangle<0$ for all magnetic  Prandtl numbers. 
Both the Reynolds stress as well as the Maxwell stress are thus transporting angular momentum outwards. As expected for negative shear,  $b_R$ and $b_\phi$ are  anticorrelated. Mainly the Maxwell stress transports the angular momentum outward but this is only true for large magnetic Prandtl numbers.


\begin{figure}
  \centerline{
 \includegraphics[width=0.7\textwidth]{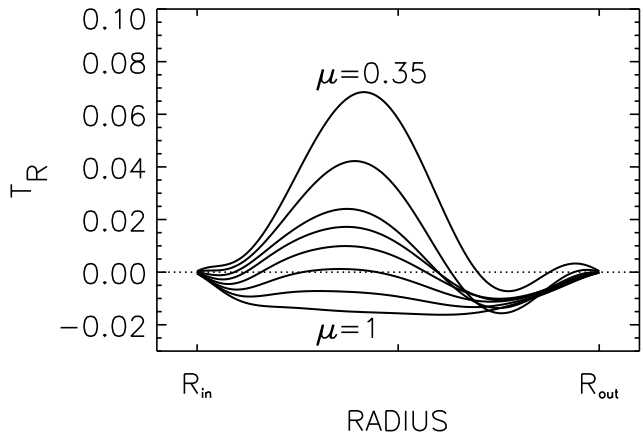}
  }
   \caption{The normalized angular momentum transport for  nonuniformly rotating
 $z$-pinches with $\mu=1$  to $\mu=0.35$  (marked).
 $\mu_B=2$, $\Ha=50$,  $m=\pm 1$, $\Pm=0.1$, $\rin=0.5$. Perfect-conducting  cylinders.}
\label{fig51}
\end{figure}

After the Boussinesq relation (\ref{Bou}) a $z$-pinch with negative shear generates positive $T_R$ for all $\Pm$. At least for $\Pm=0.1$ there must thus a transition exist between the negative values for rigid rotation and the positive values for, e.g., Kepler rotation as shown by the right panel of Fig. \ref{fig42}. The questions arise if  this transition is monotonous and for which rotation law the angular momentum flux vanishes. For the rotation laws with $\mu=1$, $\mu=0.9,...., 0.35$ the normalized values of $T_R$ are given in Fig. \ref{fig51} within the cylindrical gap for $\Ha=50$. 
 
One finds $T_R$ vanishing for $\mu\simeq 0.75$.  If (\ref{Bou}) is modified to
\beg
T_R=\Lambda \Om-\nu_{\rm T}  R\frac{\dd\Om}{\dd R}, 
\label{Bou2}
\ende
then $\nuT=\Lambda\  \dd \log R/\d\log \Om$  for $\mu=0.75$. 
In the sense of an heuristic estimate  $ \dd \log \Om/\dd \log R\simeq - 2/3$ is used so that
\beg
\nuT \simeq 0.05 \frac{E}{\Omega}
\label{Bou3}
\ende
follows with the total turbulence energy  $E$ (see Sect. \ref{Rotating}). If $E$ can be replaced by $P/\rho$ with $P$ as the turbulence pressure then Eq. (\ref{Bou3}) can be read as a confirmation of the viscosity approximation introduced by \cite{SS73}. Note that the numerical coefficient in Eq. (\ref{Bou3}) is basically smaller than unity. Though with a linear theory  the function $E=E(\Om)$ cannot be determined so that the numerical calculation of the eddy viscosity must remain open. If it should be allowed to work with the phase velocity $\omega_{\rm dr}/k$ (the ratio of   the drift frequency    and the wave number)  for the r.m.s. velocity then 
\beg
\frac{\nuT}{\nu} \simeq 0.05 \left(\frac{\omega_{\rm dr}}{k}\right)^2\ \Rey
\label{Bou4}
\ende
with $\omega_{\rm dr}/k$ in code units. The latter ratio proved to be rather smooth along the neutral line,  with a characteristic  value of  0.02 for $\Rey=~$O(10$^4$).  Hence,  $\nuT/\nu\simeq 10^{-3}\ \Rey$, in rough agreement to   nonlinear results  \citep{RG18}.
\section{Conclusions}\label{Conclusions}
We have shown that for the magnetic instability of azimuthal fields  the angular momentum transport can often be modeled by the diffusion approximation (\ref{Bou}) but not always. Exceptions exist  for  rigidly rotating tubes with  radially increasing magnetic toroidal magnetic fields where the axial background current flows in the fluid   parallel to the rotation axis. The resulting nonvanishing radial angular momentum flux excludes  the { uniform} rotation as   a solution of the MHD equation system of the rotating pinch.

For reasons of consistency our models are mostly located at the lines of neutral  stability   for given radial profiles of $U_\phi$ and $B_\phi$ where the growth rates vanish.  In one case we also proceeded along  a vertical cut in the  ($\Ha/\Rey$) plane with a fixed Hartmann number ($\Ha=50$) for Reynolds numbers smaller than $\Remax$ where the growth rates are positive. The results demonstrate the existence of normalized angular momentum fluxes even exceeding the values  $T_R(\Remax)$.

Among the fields which we probe for angular momentum transport are also those of the \C-type  
where  the magnetic field $B_\phi$ has  the same radial profile as the linear velocity $U_\phi$ of the rotation. The instability curves  of such \C-type systems coincide for small $\Pm$ in the ($\Ha/\Rey$) plane. 
Prominent examples of this particular class of MHD flows 
 are the    rigidly rotating $z$-pinch (flow and field are linearly running with $R$)  and also the  quasi-Keplerian  rotation  combined with a magnetic  field running with $1/\sqrt{R}$). For both constellations one finds finite values of the angular momentum transport. It is  positive (outward flow of angular momentum)  for the quasi-Keplerian rotation for all $\Pm$ and   for the uniformly rotating pinch but only for  large  magnetic Mach number. For slow rigid rotation and not too large $\Pm$  it is negative, i.e. the angular momentum flows  inward. 

For small magnetic Prandtl number  the contribution of the Maxwell stress  to the angular momentum flow is only small but the Reynolds stress $\langle u_R u_\phi\rangle$ is negative. For the larger $\Pm$ the transport is always outward due to the dominating negative Maxwell stress, $\langle b_R b_\phi\rangle<0$, as expected for rotation with negative shear. 
\begin{figure}
  \centerline{
 \includegraphics[width=0.49\textwidth]{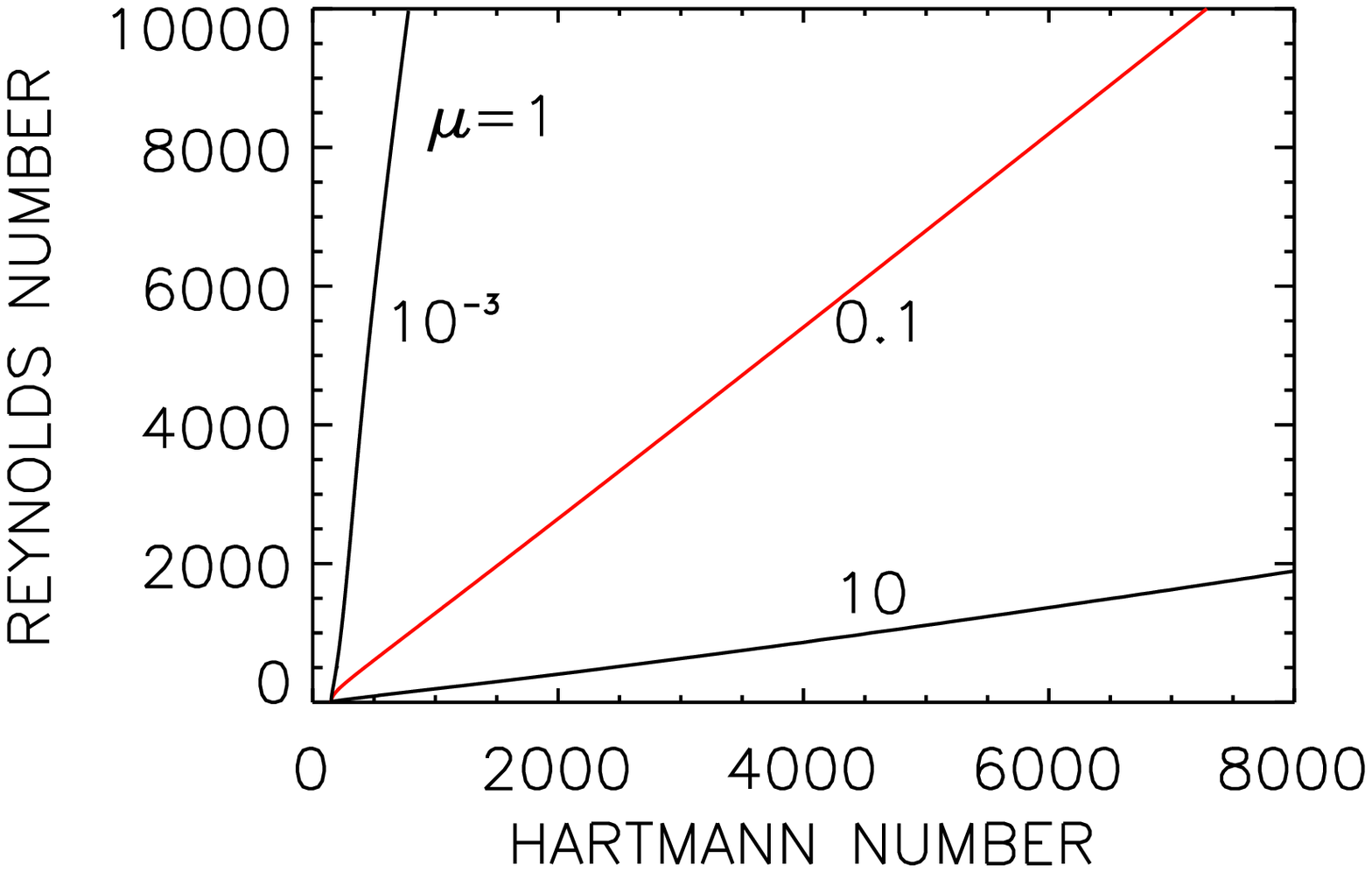}
   \includegraphics[width=0.49\textwidth]{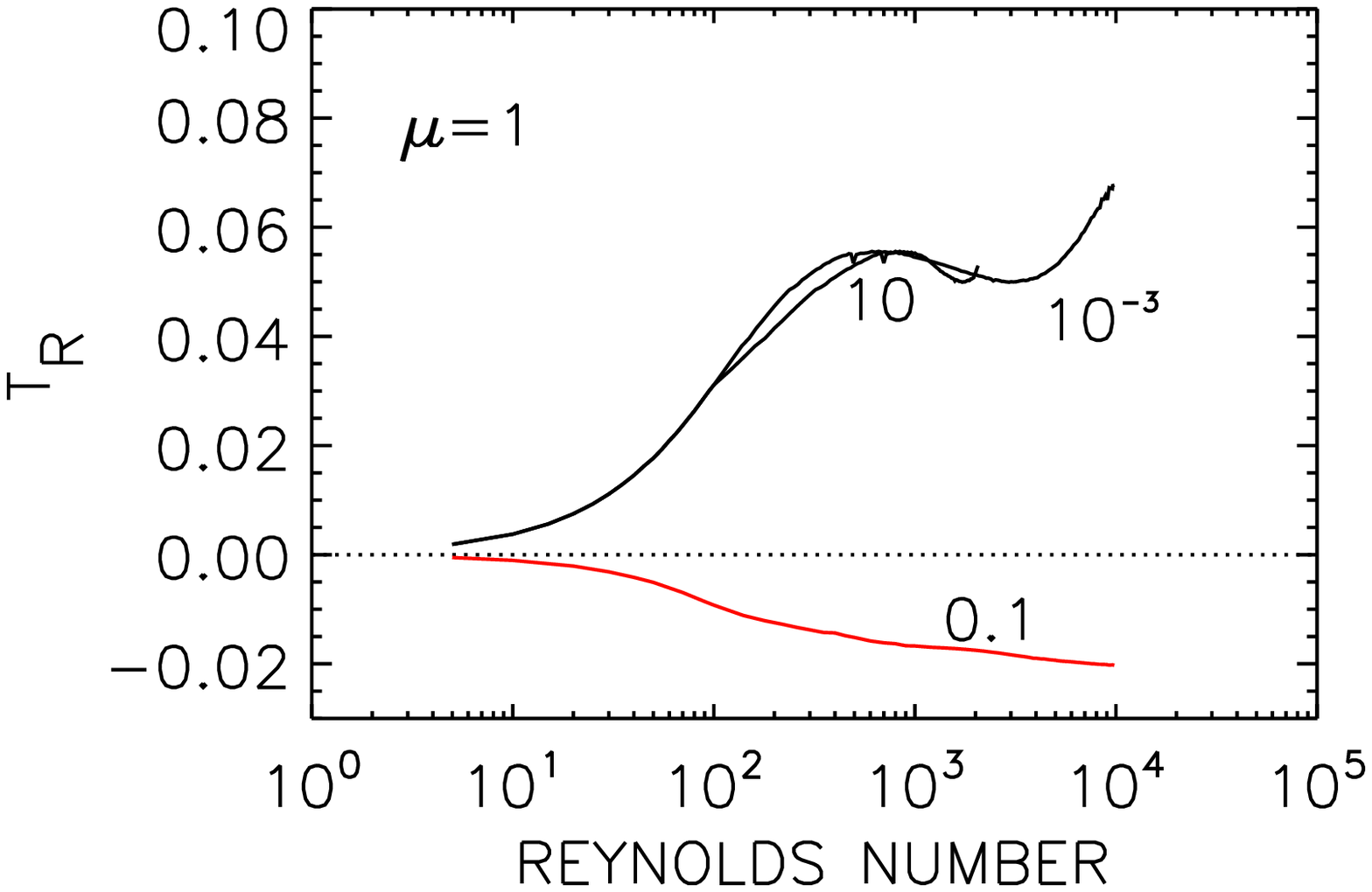}
    }
  \caption{Stability map (left) and radial angular momentum transport (right)  for uniform rotation  and quasi-uniform magnetic field for various magnetic Prandtl numbers. Flows with  Reynolds numbers  above  the lines are stable. It is $\Ha_0=150$ for $\Rey=0$. The curves are marked with their values of $\Pm$. 
   $\mu=\mu_B=1$, $m=\pm 1$, $\rin=0.5$. Perfect-conducting  cylinders.}
\label{quasiuniform}
\end{figure}

We have checked the  surprising finding  for rigid rotation also with the the quasi-uniform  field as  another prominent  magnetic field profile ($\mu_B=\mu=1$). Flow and field of this model  (which 
is  not of  the \C-type) are only unstable for $\Mm<0.5$. Again  a finite angular momentum transport exists which  again is  directed inward for  $\Pm=0.1$ (Fig.  \ref{quasiuniform}).  Obviously, the radial profile of the magnetic field does not play the decisive role for the existence of the $\Lambda$ effect of magnetic instability.

The phenomenon that   a {\em rigidly} rotating $z$-pinch for a given magnetic Mach  number transports angular momentum cannot be described by the diffusion approximation (\ref{Bou}). 
It   forms  a magnetic  counterpart  to the hydrodynamical $\Lambda $ effect in rotating anisotropic turbulences. It  means that solid-body  rotation can neither be maintained in  rotating convection zones nor   in rotating tanks  with a conducting fluid and a supercritical electric current flowing in $z$-direction. 

 It makes also sense to study the transition of the angular momentum flux   from uniform rotation (inward transport) to quasi-Keplerian rotation (outward transport). Figure \ref{fig51}  demonstrates how the anomalous angular momentum transport  disappears if the rotation law becomes more and more nonuniform. At a certain shear value the angular momentum flux  vanishes. For the $z$-pinch with $\Pm=0.1$ the transport vanishes for $\mu\simeq 0.75$ where $\Lambda$ effect and viscous transport compensate each other.  This result can be used for an approximated evaluation of the eddy viscosity. The related expression formally explains the formulation of \cite{SS73} and  numerical values (by use of  the phase velocity as the r.m.s. velocity of the instability) only slightly  differ from the results of nonlinear simulations presented earlier. 


\bibliographystyle{jpp}
\bibliography{superamri}
\end{document}